\documentclass{article}
\title{Estimating Jones and HOMFLY polynomials with One Clean Qubit}

\usepackage[bottom]{footmisc}
\usepackage{graphicx}
\usepackage{amssymb}
\usepackage{amsmath}
\usepackage{fullpage}
\usepackage{dsfont}

\author{Stephen P. Jordan\footnote{Institute for Quantum Information,
    California Institute of Technology,
    Pasadena. \texttt{sjordan@caltech.edu}} \footnote{Parts of this
    work were completed while SJ was at MIT and RIKEN.} \ and Pawel 
    Wocjan\footnote{School of Electrical Engineering and Computer
    Science, University of Central
    Florida, Orlando. \texttt{wocjan@eecs.ucf.edu}}}

\date{}

\newcommand{\captionfonts}{\small}

\makeatletter  
\long\def\@makecaption#1#2{%
  \vskip\abovecaptionskip
  \sbox\@tempboxa{{\captionfonts #1: #2}}%
  \ifdim \wd\@tempboxa >\hsize
    {\captionfonts #1: #2\par}
  \else
    \hbox to\hsize{\hfil\box\@tempboxa\hfil}%
  \fi
  \vskip\belowcaptionskip}
\makeatother   

\begin{document}
\bibliographystyle{plain}

\maketitle
\newcommand{\ud}{\mathrm{d}}
\newcommand{\bra}[1]{\langle #1|}
\newcommand{\ket}[1]{|#1\rangle}
\newcommand{\braket}[2]{\langle #1|#2\rangle}
\newcommand{\Bra}[1]{\left<#1\right|}
\newcommand{\Ket}[1]{\left|#1\right>}
\newcommand{\Braket}[2]{\left< #1 \right| #2 \right>}
\renewcommand{\th}{^\mathrm{th}}
\newcommand{\tr}{\mathrm{Tr}}
\newcommand{\id}{\mathds{1}}

\newtheorem{lemma}{Lemma}
\newtheorem{theorem}{Theorem}
\newtheorem{prop}{Proposition}
\newtheorem{definition}{Definition}

\begin{abstract}
The Jones and HOMFLY polynomials are link invariants with close
connections to quantum computing.  It was recently shown that finding
a certain approximation to the Jones polynomial of the trace closure
of a braid at the fifth root of unity is a complete problem for the
one clean qubit complexity class\cite{Shor_Jordan}. This is the class
of problems solvable in polynomial time on a quantum computer acting
on an initial state in which one qubit is pure and the rest are
maximally mixed. Here we generalize this result by showing that one
clean qubit computers can efficiently approximate the Jones and
single-variable HOMFLY polynomials of the trace closure of a braid at
\emph{any} root of unity.
\end{abstract}

\section{Introduction}
\label{intro}

\begin{figure}
\begin{center}
\includegraphics[width=0.43\textwidth]{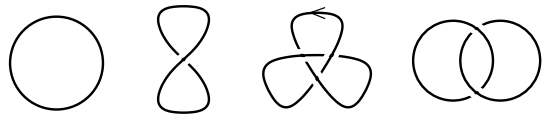}
\caption{\label{knots_and_links} Shown from left to right are the unknot,
  another representation of the unknot, an oriented trefoil knot, and
  the Hopf link. Broken lines indicate undercrossings.}
\end{center}
\end{figure}

A knot is an embedding of the circle into three dimensional
space. More generally, a link is an embedding of one or more circles
into three dimensional space. A link is said to be oriented if one of
the two possible orientations is chosen for each circle. Examples are
shown in figure \ref{knots_and_links}.

Two links are equivalent if one can be continuously deformed
into the other without cutting any strands. One of the most
fundamental tasks in the theory of links is to determine whether a
given pair of links is equivalent. Although this task appears easy in
the simple examples of figure \ref{knots_and_links}, it rapidly
becomes difficult for links of many crossings. No polynomial time
algorithm for this problem is known. Currently the best upper bound on
the complexity of the link equivalence problem is that it is contained
in NP \cite{Hass}.

Link invariants are one tool for distinguishing links. A link
invariant is some function $f$ on links such that if link $L$ is
equivalent to link $L'$ then $f(L)=f(L')$. There may exist
inequivalent links that a given link invariant fails to
distinguish. The Jones polynomial is an important link invariant that
has been very successful in distinguishing inequivalent links. It
was discovered in 1985 by Vaughan Jones \cite{Jones}. For an oriented
link $\vec{L}$ with $m$ crossings, the corresponding Jones polynomial
$V_{\vec{L}}(t)$ is a polynomial consisting of a linear combination of
integer and half-integer powers of $t$. $V_{\vec{L}}(t)$ has degree at
most $\mathcal{O}(m)$, and the coefficients in the polynomial are all
integers. That is, $V_{\vec{L}}(t) \in
\mathbb{Z}[t^{1/2},t^{-1/2}]$. The coefficients may be exponentially
large, and finding their values exactly is known to be
\#P-complete\cite{Jaeger}.

\begin{figure}
\begin{center}
\includegraphics[width=0.35\textwidth]{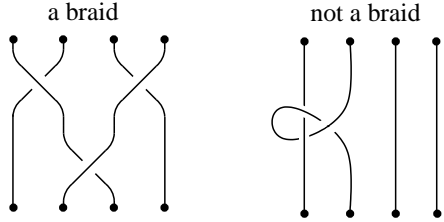}
\caption{\label{braids} On the left we have a braid of four
  strands. The strands must move steadily downwards, thus the object
  on the right is not a braid.}
\end{center}
\end{figure}

In order to formulate computational problems about links, one needs a
way to input links into a computer. One way to do this is to use the
discrete language of the braid group. A braid of $n$ strands has $n$
pegs across the top and $n$ pegs across the bottom. Each top peg is
the starting point of exactly one strand. Each bottom peg is the end
point of exactly one strand. On the way, the strands can wind around
each other in any arbitrary way, but cannot ``double back,'' as
illustrated in figure \ref{braids}. Two braids are equivalent if one
can be deformed into the other without cutting any strands. 

\begin{figure}[!htbp]
\begin{center}
\includegraphics[width=0.6\textwidth]{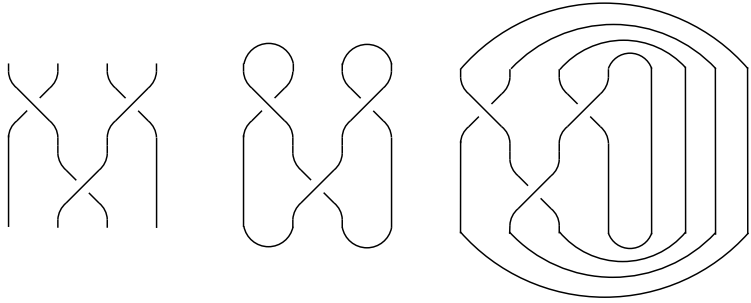}
\caption{\label{closures} Shown from left to right are a braid, its
  plat closure, and its trace closure.}
\end{center}
\end{figure}

The set of braids on $n$ strands has the structure of a group. The
group operation is concatenation of braids, as shown below. \\
\begin{center}
\includegraphics[width=0.35\textwidth]{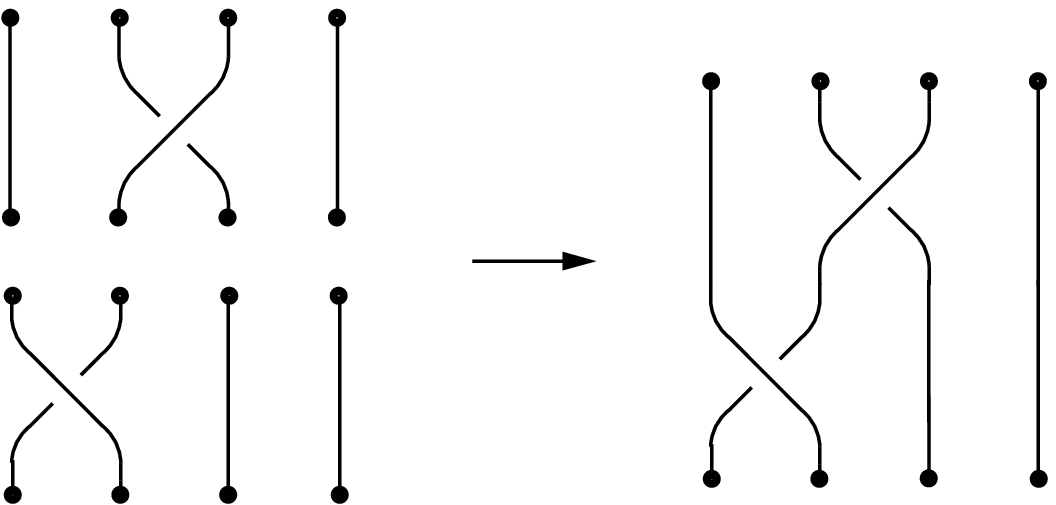}
\end{center}
The $n$-strand braid group $B_n$ is generated by the elementary
crossings $\sigma_1,\ldots,\sigma_{n-1}$ as illustrated below.\\
\begin{center}
\includegraphics[width=0.4\textwidth]{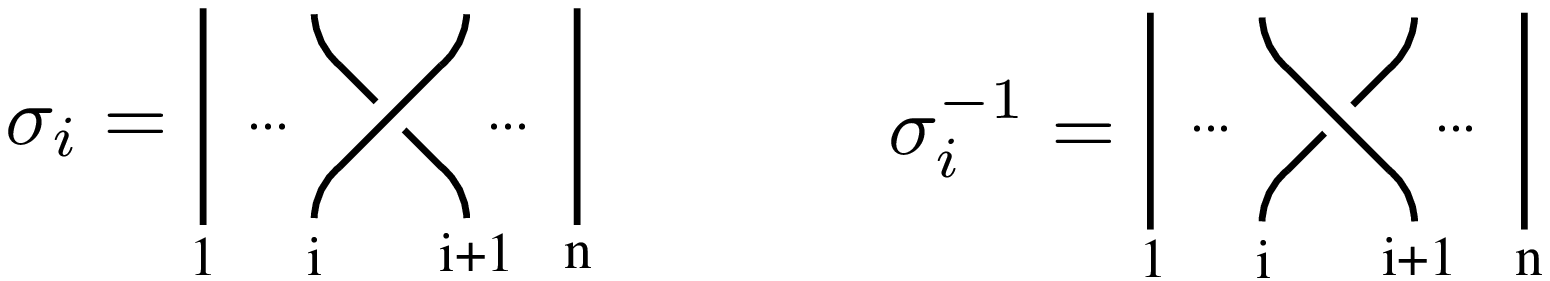}
\end{center}
For example, the braid of figure \ref{braids} is $\sigma_1^{-1}
\sigma_3 \sigma_2$. The topological equivalence of braids is
completely captured by the following two relations among the group
generators.
\begin{equation}
\begin{array}{rcll}
\sigma_i \sigma_j & = & \sigma_j \sigma_i & \textrm{ for $|i-j| \geq 2$} \\
\sigma_{i+1} \sigma_i \sigma_{i+1} & = & \sigma_i \sigma_{i+1} \sigma_i
& \textrm{ for all $i$}
\end{array}
\end{equation}
By joining the free ends of a braid, one can construct a link. Figure
\ref{closures} illustrates two ways of doing this: the plat closure
and the trace closure. Alexander's theorem states that any link can be
obtained as the trace closure of some braid. The same is true of the
plat closure\cite{Shor_Jordan}.

In addition to gaining a convenient way for inputting links into
computers, by thinking of links in terms of the braid group, we gain
an algebraic point of view on the topological problem of distinguishing
links. Jones originally formulated his polynomial in terms of certain
representations of the braid group\cite{Jones}. This original
representation-theoretic formulation is also convenient for use in
quantum computation. We'll now describe it.

Let $b$ be a braid of $n$ strands and let $b^{\mathrm{tr}}$ be the
link obtained by taking its trace closure. If each strand of the braid
is oriented downward, then an oriented link $\vec{L}$ results from
taking the trace closure. The Jones polynomial of $\vec{L}$ at $t=e^{i
  2 \pi/k}$ is\\
\begin{equation}
\label{Jones_trace}
V_{\vec{L}}(e^{i 2 \pi/k}) =
\left(-ie^{i \pi/2k}  \right)^{3 w(\vec{L})} \left(- 2 \cos(\pi/k)
\right)^{n-1} \widetilde{\tr} \left[ \rho_{n,k}(b) \right],
\end{equation}
where $w(\vec{L})$ is the ``writhe'' of $\vec{L}$. A crossing of
oriented strands of the form
\includegraphics[width=0.2in]{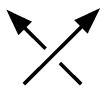}
is considered positive, and a crossing of the form
\includegraphics[width=0.2in]{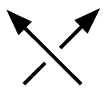} is considered
negative. $w(\vec{L})$ is equal to the number of positive
crossings minus the number of negative crossings in
$\vec{L}$. $\rho_{n,k}$ is the path model representation of the braid
group $B_n$. $\widetilde{\tr}$ is a certain weighted trace known as
the Markov trace. It is clear that the prefactor 
$\left(-ie^{i \pi/2k}  \right)^{3 w(\vec{L})} \left(- 2 \cos(\pi/k)
\right)^{n-1}$ is easy to calculate, thus the problem of evaluating
Jones polynomials polynomial-time reduces to the evaluation of the
Markov trace of the path model representation.

In 1989, Witten proved that the Jones polynomial arises as a Wilson
loop in Chern-Simons theory, thereby uncovering a
connection between topological quantum field theory and knot
invariants\cite{Witten}. In 2002, Freedman \emph{et al.} showed that
quantum computers can efficiently simulate certain topological quantum
field theories\cite{Freedman2}, and furthermore that the problem of
simulating these topological quantum field theories is
BQP-complete\cite{Freedman}. The results of Freedman \emph{et al.}
combined with that of 
Witten imply that quantum computers can efficiently estimate the
Jones polynomial of the plat closure of a braid at $t=e^{i2 \pi/5}$
and furthermore that this problem is BQP-complete. Aharonov \emph{et
  al.} subsequently generalized this result, showing that quantum
computers can efficiently estimate the Jones polynomial of the plat or
trace closure of a braid at $t=e^{i 2 \pi/k}$ for any  
$k$ \cite{Aharonov1}. In \cite{Aharonov2, Freedman3,
  Wocjan_Yard,Freedman}, the problem of estimating the Jones
polynomial of the plat closure of a braid was shown to be BQP-complete
for each $k$ other than 1,2,3,4, and 6. The problem of
estimating the Jones polynomial of the trace closure of a braid at
$t=e^{i 2 \pi/5}$ was shown in \cite{Shor_Jordan} to be complete for
the one clean qubit complexity class, called DQC1.

Whereas the Jones polynomial of the trace closure of a braid is
proportional to the Markov trace of its path model representation, the
Jones polynomial of the plat closure of a braid is proportional to a
certain matrix element of its path model representation. For $t=e^{i 2
  \pi/k}$, the path model representation is unitary. The dimension of
the representation is in general exponential in $n$. Thus the direct
classical algorithm for calculating the representation of a braid by
multiplying the matrices representing individual crossings requires
exponential time. In contrast, a quantum circuit on $n$ qubits
corresponds to an element of $U(2^n)$. By the path model
representation, a braid on $n$ strands corresponds to an exponentially
large unitary matrix, which in turn corresponds to a quantum circuit
on $\mathrm{poly}(n)$ qubits. The nontrivial achievement of
\cite{Freedman2, Aharonov1, Wocjan_Yard, Shor_Jordan} is to show that
the number of gates in the quantum circuit need only grow polynomially
with the number of crossings in the braid.

Such a correspondence between braids and quantum circuits forms the
core of the completeness proofs for Jones polynomial
problems. Estimating a matrix element of a quantum circuit to
polynomial precision is BQP-complete, and estimating the normalized
trace of a quantum circuit to polynomial precision is
DQC1-complete. Constructing the correspondence between braids and
circuits is slightly more involved in the case of DQC1-completeness
essentially because the circuit can only use logarithmically many
ancilla qubits\cite{Shor_Jordan}.

The approximations to Jones polynomials obtained by quantum computers
are additive. The Markov trace $\widetilde{\tr}(\rho_{n,k}(b))$ has
magnitude at most one. The quantum algorithm for approximating the
trace closure produces an estimate $e$ satisfying  $|e -
\widetilde{\tr}(\rho_{n,k}(b))| \leq \epsilon$ with probability
$1-\delta$ in $\mathrm{poly}(1/\epsilon, \log(1/\delta))$ time. It is
important to distinguish this from the other common type of approximation
known as a Fully Polynomial Randomized Approxation Scheme (FPRAS). An
FPRAS for a function $f$ produces an estimate $e$ satisfying 
$(1-\epsilon)f \leq e \leq (1+\epsilon)f$ with probability $1-\delta$
in time $\mathrm{poly}(1/\epsilon, \log(1/\delta))$. For many braids
$b \in B_n$, $|\widetilde{\tr}(\rho_{n,k}(b))|$ is exponentially small
compared to one. For these instances, an FPRAS is exponentially more
precise than a polynomial additive approximation.

The discovery of the Jones polynomial broke open a new field. A number
of new and powerful knot invariants related to the Jones polynomial
were soon discovered. The HOMFLY polynomial\footnote{The name HOMFLY
  stands for the names of the discoverers of this invariant: Hoste,
  Ocneanu, Millett, Freyd, Lickorish, and Yetter. Some authors prefer
  the name HOMFLYPT polynomial to recognize the contributions of
  Przytycky and Traczyk. We will use the term HOMFLY polynomial simply
  because it is more widespread.} \ is one of these. Like the Jones
polynomial, the HOMFLY polynomial is an invariant of oriented 
links. In general the HOMFLY polynomial is a polynomial in two
variables, $H_{\vec{L}}(t,x) \in \mathbb{Z}[t,t^{-1},x,x^{-1}]$. An
important special case is the single-variable HOMFLY polynomial 
\[
H_{\vec{L}}^{(r)}(q) \equiv H_{\vec{L}}(q^{r/2},q^{1/2}-q^{-1/2}),
\]
also known as the $\mathfrak{sl}_r$ invariant. As discussed in
the appendix, the Jones polynomial is
equivalent to the $r=2$ special case of the single-variable HOMFLY
polynomial. In \cite{Wocjan_Yard}, Wocjan and Yard showed that quantum
computers can efficiently approximate single-variable HOMFLY
polynomials at arbitrary roots of unity. The HOMFLY polynomial is in
turn a special case of an extremely general combinatorial object
called the Tutte polynomial. Aharonov \emph{et al.} have obtained
efficient quantum algorithms for approximating Tutte
polynomials\cite{Aharonov3}. It is not yet fully known for what range
of parameters the approximation obtained in \cite{Aharonov3} is
BQP-hard. 

The one clean qubit model was introduced in \cite{Knill} as
an idealized model of quantum computation on highly mixed states. For
example, the states manipulated in NMR experiments are typically
highly mixed. One clean qubit computers are believed to be less
powerful than standard quantum computers but still capable of solving
some problems outside of P. In the one clean qubit model one is given
an initial state consisting of one qubit in the pure state $\ket{0}$
and $n$ qubits in the maximally mixed state. In other words, the
initial density matrix is
\[
\rho = \ket{0} \bra{0} \otimes \frac{I}{2^n},
\]
where $I$ is the $2^n \times 2^n$ identity matrix. One is then allowed
to apply polynomially many quantum gates to this state, and then do a
single-qubit measurement in the computational basis. This procedure can
be repeated polynomially many times, each time starting with the same
initial state $\rho$. The set of decision problems solvable by this
procedure is called DQC1.

Here we show that one clean qubit computers can efficiently estimate
Jones and HOMFLY polynomials at arbitrary roots of unity,
generalizing the result of \cite{Shor_Jordan}. To do this we need only
two facts about one clean qubit computers. First, one clean qubit
computers can efficiently estimate the normalized trace of quantum
circuits to polynomial precision\footnote{Although we do not need
  this fact here, it is interesting to note that the decision version
  of this problem is DQC1-complete.}.  That is, we are given a
classical description of a quantum circuit on $n$ qubits with
$\mathrm{poly}(n)$ gates. This quantum circuit implements some unitary
transformation $U$ on a $2^n$-dimensional Hilbert space. The quantity
$\frac{\tr[U]}{2^n}$ is a complex number of magnitude at most one. One
clean qubit computers can produce an estimate of $T_U$ such that with
probability $1-\delta$, $\left| T_U - \frac{\tr[U]}{2^n} \right| <
\epsilon$ in time $\mathrm{poly}(1/\epsilon, \log(1/\delta))$. Second,
a computer with one clean qubit can simulate a computer with
$\mathcal{O}(\log n)$ clean qubits with polynomial overhead. Both of
these facts are discussed thoroughly in \cite{Shor_Jordan}. For
additional information about one clean qubit computers we refer the
interested reader to \cite{Knill, qwgt, decay1, decay2, Ambainis,
  Shor_Jordan, Datta}.

\section{Path Model Representation of $B_n$}
\label{pathmodel}
As discussed in section \ref{intro} and reference
\cite{Aharonov1}, the problem of estimating the Jones polynomial of
the trace closure of a braid reduces to the problem of estimating 
the Markov trace of the braid's path model representation. 
In this section we present the path model representation of the braid
group $B_n$, and the Markov trace of this representation.

Let $\Omega_{n,k}$ be the set of paths of $n$ steps on a ladder of
$k-1$ rungs that start at the bottom. For example
\[
\Omega_{4,4} = \left\{ 
\includegraphics[trim=0in 0.095in 0in -0.06in, width=0.5in]{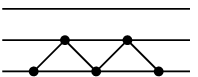},
\includegraphics[trim=0in 0.095in 0in -0.06in, width=0.5in]{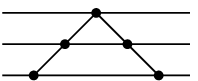},
\includegraphics[trim=0in 0.095in 0in -0.06in, width=0.5in]{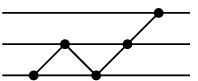},
\includegraphics[trim=0in 0.095in 0in -0.06in, width=0.5in]{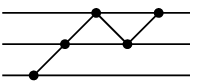}
\right\}.
\]
Let $\mathcal{V}_{n,k}$ be the formal span of $\Omega_{n,k}$ and let
$\mathcal{V}^\dag_{n,k}$ be its dual. For example
\begin{eqnarray*}
\mathcal{V}_{4,4} & = & \mathrm{span} \left\{
\Ket{\includegraphics[trim=0in 0.095in 0in -0.06in, width=0.5in]{bp1.eps}},
\Ket{\includegraphics[trim=0in 0.095in 0in -0.06in, width=0.5in]{bp2.eps}},
\Ket{\includegraphics[trim=0in 0.095in 0in -0.06in, width=0.5in]{bp3.eps}},
\Ket{\includegraphics[trim=0in 0.095in 0in -0.06in, width=0.5in]{bp4.eps}}
\right\} \\
\mathcal{V}^\dag_{4,4} & = & \mathrm{span} \left\{
\Bra{\includegraphics[trim=0in 0.095in 0in -0.06in, width=0.5in]{bp1.eps}},
\Bra{\includegraphics[trim=0in 0.095in 0in -0.06in, width=0.5in]{bp2.eps}},
\Bra{\includegraphics[trim=0in 0.095in 0in -0.06in, width=0.5in]{bp3.eps}},
\Bra{\includegraphics[trim=0in 0.095in 0in -0.06in, width=0.5in]{bp4.eps}}
\right\}
\end{eqnarray*}
where for any $p,q \in \Omega_{n,k}$:
\[
\braket{p}{q}=\delta_{p,q}.
\]
For any $n,k \in \mathbb{N}$ the path model representation
$\rho_{n,k}$ is a homomorphism from $B_n$, the $n$-strand braid group,
to $U(\mathcal{V}_{n,k})$, the group of unitary transformations on
$\mathcal{V}_{n,k}$.

Let $\sigma_i$ denote the crossing of strands $i$ and $i+1$:
\[
\includegraphics[width=1.0in]{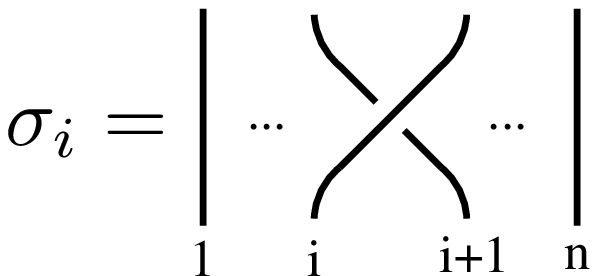}
\]
$\rho_{n,k}(\sigma_i)$ acts only on steps $i$ and $i+1$ of paths in
$\Omega_{n,k}$ leaving the other steps unchanged. Specifically,
\[
\includegraphics[width=0.8\textwidth]{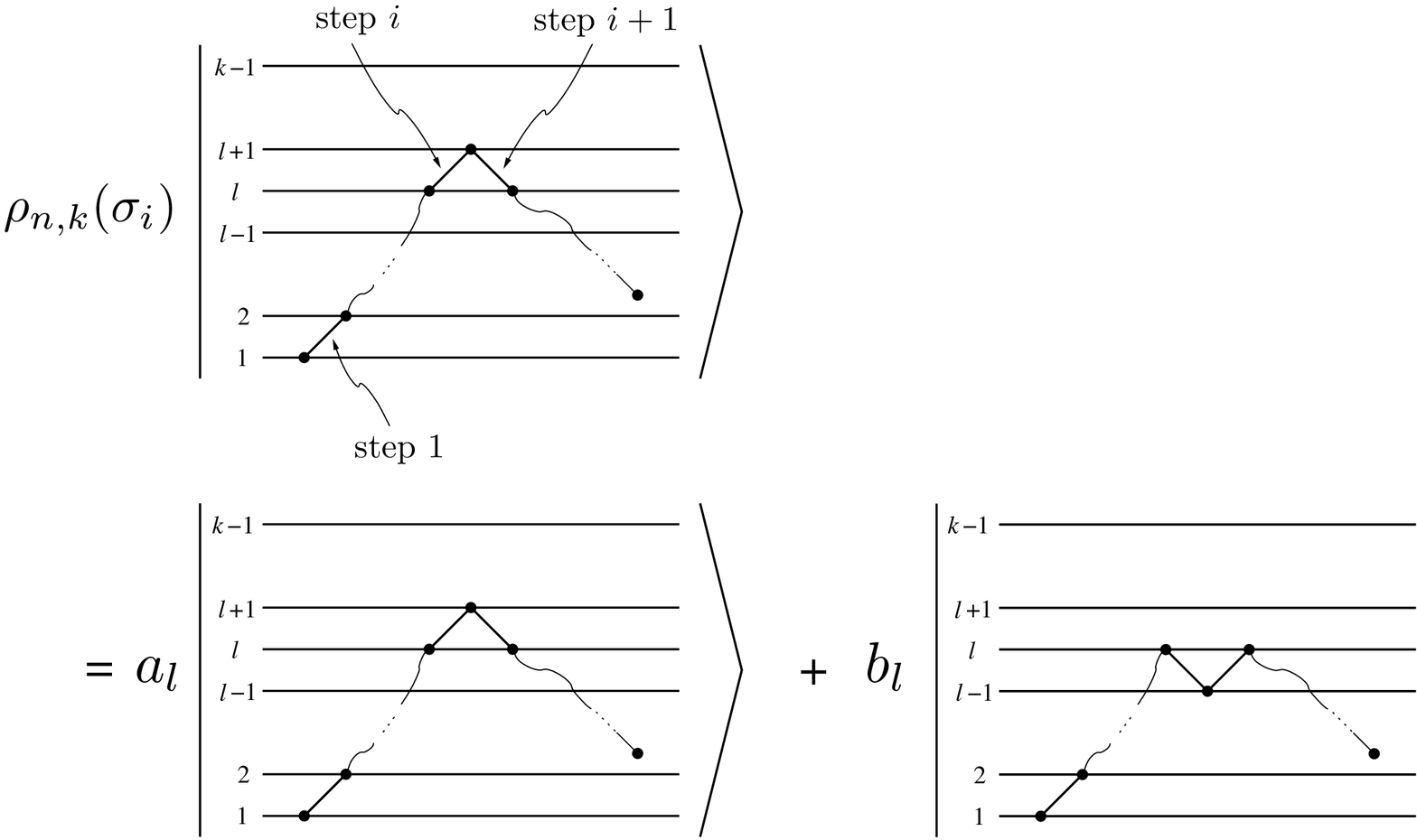}
\]
And similarly,
\begin{eqnarray*}
\rho_{n,k}(\sigma_i) 
\Ket{\includegraphics[trim=0in 0.2in 0in -0.06in,width=0.75in]{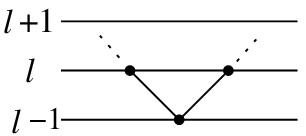}} & = &
c_l \Ket{\includegraphics[trim=0in 0.2in 0in -0.06in,width=0.75in]{du.eps}}
+ d_l \Ket{\includegraphics[trim=0in 0.2in 0in -0.06in,width=0.75in]{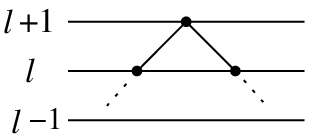}} \\
\rho_{n,k}(\sigma_i) 
\Ket{\includegraphics[trim=0in 0.2in 0in -0.06in,width=0.75in]{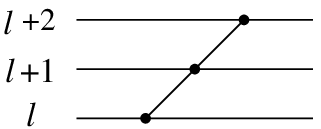}} & = &
e_l \Ket{\includegraphics[trim=0in 0.2in 0in -0.06in,width=0.75in]{uu.eps}} \\
\rho_{n,k}(\sigma_i) 
\Ket{\includegraphics[trim=0in 0.2in 0in -0.06in,width=0.75in]{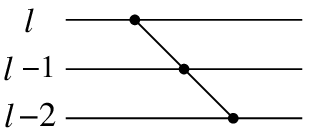}} & = &
f_l \Ket{\includegraphics[trim=0in 0.2in 0in -0.06in,width=0.75in]{dd.eps}}
\end{eqnarray*}
where:
\[
\begin{array}{cclcccl}
A & = & i e^{-i \pi/2k} & \quad & \lambda_l & = & \sin \left(
\frac{\pi l}{k} \right) \vspace{5pt}\\
e_l & = & A^{-1} & \quad & f_l & = & A^{-1} \vspace{5pt} \\
a_l & = & A^{-1} + A \frac{\lambda_{l+1}}{\lambda_l} & \quad & c_l & = &
A^{-1} + A \frac{\lambda_{l-1}}{\lambda_l} \vspace{5pt} \\
b_l & = & d_l \ = \ A \frac{\sqrt{\lambda_{l+1}
    \lambda_{l-1}}}{\lambda_l} & & & &
\end{array}
\]
These rules completely define the representation $\rho_{n,k}$.

Let $\Omega_{n,k,h}$ be the set of paths in $\Omega_{n,k}$ that end on
rung $h$. Let $\mathcal{V}_{n,k,h}$ be the corresponding
$|\Omega_{n,k,h}|$-dimensional vector space. $\rho_{n,k}(\sigma_i)$
leaves $h$ unchanged for all $i$, as one can see from the preceding
rules. Thus, for each $h$, these rules define a representation
$\rho_{n,k,h}:B_n \to U(\mathcal{V}_{n,k,h})$. $\rho_{n,k}$ is the
direct sum of these.
\[
\rho_{n,k}(\sigma_i) = \bigoplus_{h=1}^{k-1} \rho_{n,k,h}(\sigma_i)
\]
The Markov trace of the representation $\rho_{n,k}$ is given by:
\begin{equation}
\label{Markov_Jones}
\widetilde{\tr}(\rho_{n,k}(b)) = \frac{1}{\sum_{h=1}^{k-1} \lambda_h |
  \Omega_{n,k,h} |} \sum_{h=1}^{k-1} \tr \left[ \rho_{n,k,h}(b)
  \right] \lambda_h
\end{equation}
where $\tr$ is the ordinary matrix trace, and $\lambda_h = \sin
\left( \frac{\pi h}{k} \right)$. 

In section \ref{implementation} we show how to estimate the normalized
trace
\begin{equation}
\label{normalized}
\frac{1}{|\Omega_{n,k,h}|}\tr[\rho_{n,k,h}(b)]
\end{equation}
on a one clean qubit computer for each $h$. Given the ability to do
this, it is a simple matter to obtain the full Markov
trace. By equation \ref{Markov_Jones} we see that we can obtain the
Markov trace by classically sampling $h$ according to the distribution
\[
p(h) = \frac{\lambda_h |\Omega_{n,k,h}|}
{\sum_{h=1}^{k-1} \lambda_h |\Omega_{n,k,h}|}.
\]
For each $h$ obtained by sampling from this distribution, we use a one
clean qubit computer to estimate the corresponding normalized trace of
equation \ref{normalized}. By construction, the average obtained by
this sampling procedure will converge to the Markov
trace. By taking polynomially many samples, one can obtain the
Markov trace to polynomial precision. The probability
distribution $p(h)$ is easy to sample from because $h$ can take on
only $k-1$ different values, and each $p(h)$ is furthermore easy to
compute.

To estimate the normalized trace (eq. \ref{normalized}) on a one clean
qubit computer, we introduce an encoding $\eta_h$ from bits to paths
\[
\eta_h: \{ 0,1 \}^{n\beta} \to \Omega_{n,k,h},
\]
where $\beta$ is a parameter whose value we determine in section
\ref{encoding}.  $\eta_h$ is a non-injective map. However, the number
of different bitstrings that map to a given path is approximately the
same for all paths. That is, $| \eta_h^{-1}(\omega)| \simeq
2^{n\beta}/|\Omega_{n,k,h}|$ for all $\omega \in \Omega_{n,k,h}$ where 
$|\eta^{-1}_h(\omega)|$ is the number of bitstrings in $\{0,1\}^{n\beta}$
that get mapped to $\omega$.

For any $b \in B_n$ with $\mathrm{poly}(n)$ crossings we obtain a
quantum circuit $U_{\mathrm{pm}}(b)$ of $\mathrm{poly}(n)$ gates such
that for almost all $x,y \in \{0,1\}^{n\beta}$,
\begin{equation}
\label{condition}
\bra{x} U_{\mathrm{pm}}(b) \ket{y} \simeq \bra{\eta_h(x)} \rho_{n,k,h}(b)
\ket{\eta_h(y)}.
\end{equation}
In other words, this quantum circuit implements the path model
representation of braid $b$. Thus
\[
\frac{1}{2^{n\beta}} \tr [U_{\mathrm{pm}}(b)] = \frac{1}{2^{n\beta}}
  \sum_{x \in \{0,1\}^{n\beta}} \bra{x} U_{\mathrm{pm}}(b) \ket{x}
\]
\[
\simeq \frac{1}{2^{n\beta}} \sum_{x \in \{0,1\}^{n\beta}}
\bra{\eta_h(x)} \rho_{n,k,h}(b) \ket{\eta_h(x)}
\]
\[
= \frac{1}{2^{n\beta}} \sum_{\omega \in \Omega_{n,k,h}} |
\eta^{-1}_h(\omega) | \bra{\omega} \rho_{n,k,h}(b) \ket{\omega}
\]
\[
\simeq \frac{1}{|\Omega_{n,k,h}|} \sum_{\omega \in \Omega_{n,k,h}}
\bra{\omega} \rho_{n,k,h}(b) \ket{\omega}
\]
\[
= \frac{1}{|\Omega_{n,k,h}|} \tr \left[ \rho_{n,k,h}(b) \right].
\]
With such an encoding we are able to use $n\beta$ maximally mixed
qubits to obtain a uniformly weighted trace over all paths in
$\Omega_{n,k,h}$.

\section{Encoding Paths as Bitstrings}
\label{encoding}

To describe $\eta_h$ we imagine a randomized classical algorithm
which uses $n\beta$ random bits to produce an element of $\Omega_{n,k,h}$
approximately uniformly at random. Such an algorithm corresponds to a
map from $\{0,1\}^{n\beta}$ to $\Omega_{n,k,h}$, and the fact that the
probability distribution over paths is approximately uniform ensures
that
\[
| \eta_h^{-1}(\omega)| \simeq \frac{2^{n\beta}}{|\Omega_{n,k,h}|}
\]
for all $\omega \in \Omega_{n,k,h}$. This algorithm is not to be run,
but is rather a conceptual tool for the design of the encoding $\eta_h$.

\begin{figure}
\begin{center}
\includegraphics[width=0.5\textwidth]{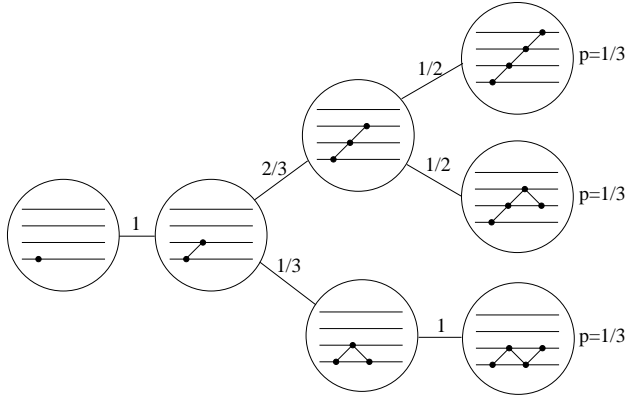}
\caption{\label{probtree} Here the transition probabilities are
  illustrated in the randomized algorithm for producing paths of three
  steps on a ladder of four rungs. Using the rule of equation
  \ref{transp}, the final probabilities come out uniform.}
\end{center}
\end{figure}

The algorithm works by starting at the bottom rung, and adding
steps one by one until a path of $n$ steps is obtained. Let
$Q^k_n(a,a')$ be the number of paths of $n$ steps on a ladder of $k-1$
rungs which start at rung $a$ and end at rung $a'$. Suppose the current
path has $t$ steps and ends at rung $a$. There are $Q^k_{n-t}(a+1,h)$
completions of this path in which step $t+1$ is upward and
$Q^k_{n-t}(a-1,h)$ completions in which step $t+1$ is downward. Thus
the algorithm chooses step $t+1$ to be upward with probability
\begin{equation}
\label{transp}
p_{\mathrm{up}}(a,t) =
\frac{Q^k_{n-t}(a+1,h)}{Q^k_{n-t}(a+1,h)+Q^k_{n-t}(a-1,h)}.
\end{equation}
(To cover the cases $a=1$ and $a = k-1$ we define $Q^k_{k,h} =
Q^k_{0,h} = 0$.) By choosing each step according to equation
\ref{transp}, one obtains at the end a uniform distribution over
$\Omega_{n,k,h}$. This is illustrated in figure \ref{probtree}. 
To generate a path of $n$ steps, we use $n$ registers of $\beta$
random bits. We think of the registers as encoding numbers
$r_1,\ldots,r_n$ in the range $0,1,\ldots,2^\beta-1$. The $t\th$ step
is chosen to be up if and only if 
\begin{equation}
\label{roundp}
r_t < \lceil p_{\mathrm{up}}(a,t) 2^\beta \rceil.
\end{equation}
Note that if the path has reached the top or bottom rung
$p_{\mathrm{up}}$ can equal 1 or 0.

If $p_{\mathrm{up}}(a,t)$ were implemented exactly then the paths would
be produced with exactly uniform probability. Because of the rounding
shown in equation \ref{roundp}, each $p_{\mathrm{up}}(a,t)$ is only
accurate to within $\pm 2^{-\beta}$. Correspondingly, the number of
bitstrings that get mapped by $\eta_h$ to a given path is not
precisely the same for all paths. This introduces an error into the
estimate of the normalized trace given by
\[
E_{\mathrm{round}} =
\left| \frac{1}{|\Omega_{n,k,h}|} \sum_{\omega \in \Omega_{n,k,h}}
\bra{\omega} \rho_{n,k,h}(b) \ket{\omega} - \frac{1}{2^{n\beta}}
\sum_{x \in \{0,1\}^{n\beta}} \bra{\eta_h(x)} \rho_{n,k,h}(b)
\ket{\eta_h(x)} \right|
\]
By the definition of $\eta_h$ this is
\[
E_{\mathrm{round}} = \left| \sum_{\omega \in \Omega_{n,k,h}}
p_{\mathrm{uni}}(\omega) \bra{\omega} \rho_{n,k,h}(b) \ket{\omega} -
\sum_{\omega \in \Omega_{n,k,h}} \widetilde{p}(\omega) \bra{\omega}
\rho_{n,k,h}(b) \ket{\omega} \right|,
\]
where $\widetilde{p}(\omega)$ is the distribution over paths produced
by the classical algorithm using $\beta$ bits of precision, and
$p_{\mathrm{uni}}(\omega) = \frac{1}{|\Omega_{n,k,h}|}$ is the uniform
distribution. By the triangle inequality,
\[
E_{\mathrm{round}} \leq \sum_{\omega \in \Omega_{n,k,h}} \left| \left(
p_{\mathrm{uni}}(\omega) - \widetilde{p}(\omega) \right) \bra{\omega}
\rho_{n,k,h}(b) \ket{\omega} \right|.
\]
Because $\rho_{n,k,h}$ is unitary this gives us
\begin{eqnarray}
\label{eround}
E_{\mathrm{round}} & \leq & \sum_{\omega \in \Omega_{n,k,h}} \left| p_{\mathrm{uni}} -
\widetilde{p}(\omega) \right| \nonumber \\
& = & \left\| p_{\mathrm{uni}} - \widetilde{p} \right\|_1.
\end{eqnarray}
Here we are thinking of probability distributions as vectors and
measuring their distance using the 1-norm.

For the purpose of estimating Jones polynomials in DQC1, one wants to
estimate the normalized trace to polynomial precision. Thus it
suffices to have
\begin{equation}
\label{onenorm}
\| p_{\mathrm{uni}} - \widetilde{p} \|_1 = \mathcal{O} \left(
\frac{1}{\mathrm{poly}(n)} \right).
\end{equation}
As proven below, to satisfy the condition \ref{onenorm}, it is
sufficient to implement each $p_{\mathrm{up}}$ to polynomial
precision. Thus it is sufficient to choose 
$\beta = \mathcal{O}(\log n)$.

Let
\[
\Omega_{\leq n, k} = \bigcup_{t=0}^n \Omega_{t,k}.
\]
As shown in figure \ref{probtree}, our classical probabilistic
algorithm can be thought of as a Markov process on $\Omega_{\leq n,
k}$. Each element in $\Omega_{t,k}$ probabilistically transitions to
one of two possible elements in $\Omega_{t+1,k}$ with probabilities
$p_{\mathrm{up}}(a,t)$ and $1-p_{\mathrm{up}}(a,t)$. Hence we can
define a $|\Omega_{\leq n, k}|$-dimensional stochastic matrix $M$
representing our idealized algorithm. Each row contains at most two
nonzero entries which are $p_{\mathrm{up}}(a-1,t-1)$ and
$1-p_{\mathrm{up}}(a+1,t-1)$. The initial probability distribution
$p_0$ on $\Omega_{\leq n, k}$ has probability one on the zero step
path:
\[
\includegraphics[width=0.5in]{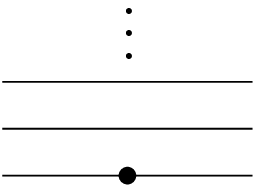}.
\]
After choosing $t$ steps, the probability distribution is
\[
p_t = M^t p_0
\]
which has support only on paths of $t$ steps. We define
$\widetilde{M}$ analogously to $M$, except that instead of
$p_{\mathrm{up}}(a,t)$ and $1-p_{\mathrm{up}}(a,t)$ the entries
represent the actual transition probabilities obtained using $\beta$ bits
of precision. Thus, in each row, $M$ and $\widetilde{M}$ have at
most two nonzero entries (at the same places) and these entries differ
by at most $\epsilon \equiv 2^{-\beta}$. Thus, in each row,
$\Delta \equiv (\widetilde{M}-M)$ has only two nonzero entries, each
of magnitude bounded by $\epsilon$. Hence for any probability
distribution $p$,
\begin{eqnarray}
\| (\widetilde{M} - M) p \|_1 & = & \sum_i \left| \sum_j
\Delta_{ij} p_j \right| \nonumber \\
& \leq & \sum_j \left[ \sum_i \left| \Delta_{ij} \right|
  \right] p_j \nonumber \\
& \leq & \sum_j 2 \epsilon p_j \nonumber \\
& = & 2 \epsilon. \label{mbound}
\end{eqnarray}
Let $\widetilde{p}_t$ be the probability distribution obtained on
$t$-step paths by the actual algorithm and let $p_t$ be that obtained
by the idealized algorithm. Further, let $E_t = \| \widetilde{p}_t -
p_t \|_1$.
\[
\begin{array}{cclcccl}
\widetilde{p}_0 & = & p_0 & \quad & E_0 & = & 0 \vspace{5pt} \\
\widetilde{p}_t & = & \widetilde{M}^t p_0 & \quad & p_t & = & M^t p_0
\end{array}  
\]
So:
\begin{eqnarray*}
E_{t+1} & = & \| \widetilde{p}_{t+1} - p_{t+1} \|_1 \\
        & = & \| \widetilde{M} \widetilde{p}_t - M p_t \|_1 \\
        & = & \| \widetilde{M} \widetilde{p}_t - \widetilde{M} p_t +
        \widetilde{M} p_t - M p_t \|_1 \\
\end{eqnarray*}
and by the triangle inequality
\begin{eqnarray*}
& \leq & \| \widetilde{M} \widetilde{p}_t - \widetilde{M} p_t\|_1
+ \| \widetilde{M} p_t - M p_t \|_1 \\
& = & \| \widetilde{M} (\widetilde{p}_t - p_t ) \|_1 + \| ( \widetilde{M}
- M) p_t \|_1.
\end{eqnarray*}
$\widetilde{M}$ is a stochastic matrix and therefore $\| \widetilde{M}
\vec{x} \|_1 \leq \| \vec{x} \|_1$ for any $\vec{x}$. Thus
\begin{eqnarray*}
E_{t+1} & \leq & \| \widetilde{p}_t-p_t \|_1 + \| (\widetilde{M}-M)
p_t \|_1 \\
& = & E_t + \| ( \widetilde{M} - M) p_t \|_1 \\
& \leq & E_t + 2 \epsilon
\end{eqnarray*}
by equation \ref{mbound}. Since $E_0 = 0$, the final error is bounded
by
\[
E_n \leq 2 n \epsilon  = 2n 2^{-\beta}.
\]
$E_n$ is exactly the expression $\|\tilde{p} - p_{\mathrm{uni}}\|_1$
appearing in equation \ref{eround}, thus choosing $\beta =
\mathcal{O}(\log n)$ suffices to make $E_{\mathrm{round}}$
polynomially small.

\section{Algorithm for Jones Polynomials}
\label{implementation}

With the encoding $\eta_h$ in place, the remaining task is to
efficiently implement $U_{\mathrm{pm}}(b)$ with a quantum circuit, as
described in equation \ref{condition}. To do this, it suffices to
efficiently implement $U_{\mathrm{pm}}(\sigma_t)$ for each crossing
$\sigma_t$. Then, to represent any $m$-crossing braid $\sigma_{t_1}
\sigma_{t_2} \ldots \sigma_{t_m}$ we can concatenate the corresponding
quantum circuits to obtain $U_{\mathrm{pm}}(\sigma_{t_1})
U_{\mathrm{pm}}(\sigma_{t_2}) \ldots U_{\mathrm{pm}}(\sigma_{t_m})$.

As discussed in section \ref{pathmodel}, $\rho_{n,k,h}(\sigma_t)$
transforms only steps $t$ and $t+1$ in any path. Hence
$U_{\mathrm{pm}}(\sigma_t)$ 
transforms only registers $t$ and $t+1$ of $\beta$ qubits each in any
encoded path. However, the transformation on these two steps depends on
the rung $l$ on which they start. Each register encodes whether a
given step is up or down. Thus, $l$ is encoded in the preceding $t-1$
registers. The number of rungs is fixed at $k-1$, thus only a constant
number of ancilla qubits ($\lceil \log_2 (k-1) \rceil$) are needed to
store $l$. As discussed in section \ref{intro},
up to logarithmically many clean ancilla qubits can be simulated on a one
clean qubit computer. We will now describe how to efficiently compute
$l$ into a register of $\mathcal{O}(1)$ clean ancilla qubits using
reversible computation. (We assume $k$ is constant, unlike $n$.)

To do this, we start by precomputing the cutoffs $\lceil 2^\beta
p_{\mathrm{up}}(a,t) \rceil$ for all $1 \leq a \leq k-1$ and $0 \leq t
\leq n$ on a standard classical computer. To store these numbers
requires $nk\beta$ bits, which for any fixed $k$ is of order $n \log
n$. By equation \ref{transp}, we can compute these cutoffs by counting
the number of paths of given length $m \leq n$ that begin and end on
given rungs.

Let $A$ be the adjacency matrix of the line graph of $k-1$ nodes
$\left( \includegraphics[width=1in]{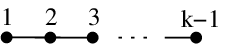} \right)$.
\[
A = \left[ \begin{array}{cccccc}
0 & 1 &   &   &   &   \\
1 & 0 & 1 &   &   &   \\
  & 1 & 0 & 1 &   &   \\
 & & & \ddots &   &   \\
  &   &   & 1 & 0 & 1 \\
  &   &   &   & 1 & 0 
\end{array} \right]
\]
Then, the number of paths of length $s$ from rung $a$ to rung $h$ is
the $a,h$ matrix element of $A^s$. This can clearly be computed in
$\mathrm{poly}(s,k)$ time.

Suppose we have one register of $c = \lceil \log_2(k-1) \rceil$ qubits
containing $l_i$, the rung of step $i$, and one register of $c$ qubits
initialized to zero in which we wish to write $l_{i+1}$. $l_{i+1}$ is
simply set to $l_i + 1$ or $l_i-1$ depending on whether the $(i+1)\th$
$\beta$-qubit register in the encoding contains a number less or
greater than the corresponding 
cutoff $\lceil 2^\beta p_{\mathrm{up}}(l_i,i) \rceil$. Comparing two
numbers to see which is bigger can be done reversibly using
logarithmically many ancillas, and the same is true for adding or
subtracting 1 to a number. (See section 4 of \cite{Shor_Jordan} for a
summary of the literature on reversible arithmetic with limited
ancillas.) Since the cutoffs are hardcoded they do not need to be
computed reversibly at all. Thus this whole process is doable in
DQC1. One then uncomputes $l_i$ and repeats this process until $l_t$
is obtained for the desired $t$. Thus starting with $l_0=1$, one can
efficiently produce a register of qubits containing $l_t$.

To implement $U_{\mathrm{pm}}(\sigma_t)$ we first compute $l_t$
and then use a unitary $U_\sigma$ that acts on three registers: the
$t\th$ and $(t+1)\th$ registers of $\beta$ qubits from the encoding
and the register containing $l_t$. $U_\sigma$ does not affect the
$l_t$ register. Rather, $l_t$ controls what operation gets applied to
the other two registers, as specified by the path model
representation. Thus, after applying $U_\sigma$, one can uncompute
$l_t$.

$U_\sigma$ acts on $2\beta + \lceil \log_2 (k-1) \rceil = \mathcal{O}(\log
n)$ qubits. By general techniques it is possible to implement
arbitrary unitary transformations on $\mathcal{O}(\log n)$ qubits
using $\mathrm{poly}(n)$ gates\cite{Nielsen_Chuang_twolevel}. Thus
efficiency is not a concern. We just need to construct a concrete
unitary implementation of $U_\sigma$ that gives the correct
transformation on the encoded paths as specified by the path model
representation.

If the encoded path is
\[
\includegraphics[width=1.4in]{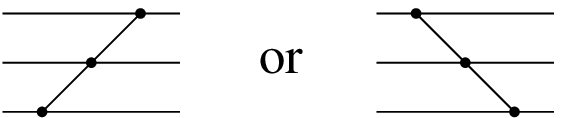}
\]
then, by the path model representation, we merely need to apply a
phase shift of $A^{-1}$. If the encoded path is
\[
\includegraphics[width=1.4in]{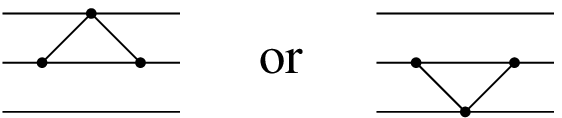}
\]
then we must unitarily transform to some linear combination of the
encodings of these two paths.

There are $2^{2\beta}$ possible values for the bits contained in the
relevant two registers. Suppose that the number of these bitstrings
that encode $\includegraphics[width=0.25in]{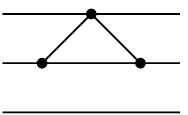}$  is equal to the
number of bitstrings that encode
$\includegraphics[width=0.25in]{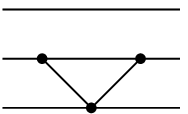}$. (As we shall see, these
two numbers are equal up to rounding.) Let's call this number
$d$. Then, we can use the labels:
\[
\Ket{\includegraphics[width=0.25in]{udsmall.eps},1},
\Ket{\includegraphics[width=0.25in]{udsmall.eps},2},
\ldots,
\Ket{\includegraphics[width=0.25in]{udsmall.eps},d},
\]
for the bitstrings that encode
$\includegraphics[width=0.25in]{udsmall.eps}$, and
\[
\Ket{\includegraphics[width=0.25in]{dusmall.eps},1},
\Ket{\includegraphics[width=0.25in]{dusmall.eps},2},
\ldots,
\Ket{\includegraphics[width=0.25in]{dusmall.eps},d},
\]
for the bitstrings that encode
$\includegraphics[width=0.25in]{dusmall.eps}$. Therefore, for each $j
\in \{1,\ldots,d\}$, $U_\sigma$ is
\begin{eqnarray}
\label{jintro}
U_\sigma \ket{l} \Ket{\includegraphics[width=0.25in]{udsmall.eps},j} &
= &
\ket{l} \left( a_l \Ket{\includegraphics[width=0.25in]{udsmall.eps},j}
+ b_l \Ket{\includegraphics[width=0.25in]{dusmall.eps},j} \right)
\nonumber \\
U_\sigma \ket{l} \Ket{\includegraphics[width=0.25in]{dusmall.eps},j} &
= &
\ket{l} \left( c_l \Ket{\includegraphics[width=0.25in]{dusmall.eps},j}
+ d_l \Ket{\includegraphics[width=0.25in]{udsmall.eps},j} \right)
\end{eqnarray}
in accordance with section \ref{pathmodel}. This is unitary and
satisfies equation \ref{condition}.

Looking in more detail at $\eta_h$ we can specify concretely the
labelling. We can think of the contents of the $t\th$ and $(t+1)\th$
registers as specifying two numbers $r_t, r_{t+1} \in
\{0,1,\ldots,2^\beta-1 \}$. Correspondingly we have the cutoffs
\[
C_t^l = \lceil 2^\beta p_{\mathrm{up}}(l,t) \rceil
\]
\[
C_{t+1}^{l+1} = \lceil 2^\beta p_{\mathrm{up}}(l+1,t+1) \rceil
\]
\[
C_{t+1}^{l-1} = \lceil 2^\beta p_{\mathrm{up}}(l-1,t+1) \rceil
\]
The bitstrings with $r_t < C_t^l$, $r_{t+1} \geq C_{t+1}^{l+1}$ encode 
$\includegraphics[width=0.25in]{udsmall.eps}$, and the bitstrings with
$r_t \geq C_t^l$, $r_{t+1} < C_{t+1}^{l+1}$ encode
$\includegraphics[width=0.25in]{dusmall.eps}$. By the definition of
$p_{\mathrm{up}}(a,t)$, the probability of hopping up and then down is
the same as the probability of hopping down and then up. This is
because both processes end on the same rung, thus each of these paths
have the same number of completions ending at height $h$ after $n-t$
additional steps. Hence, up to rounding, the number of bitstrings from
the $2^{2\beta}$ possibilities that encode
$\includegraphics[width=0.25in]{udsmall.eps}$ is the same as the
number of bitstrings that encode
$\includegraphics[width=0.25in]{dusmall.eps}$. We can choose the label
$j$ from equation \ref{jintro} to be:
\begin{eqnarray}
\mathrm{for} \ \ \includegraphics[width=0.25in]{udsmall.eps}: 
& & j= C_{t+1}^{l+1}r_t + (r_{t+1}-C_{t+1}^{l+1}) \nonumber \\
\label{placevalue}
\mathrm{for} \ \ \includegraphics[width=0.25in]{dusmall.eps}:
& & j = C_{t+1}^{l-1}(r_t-C_t^l) + r_{t+1}
\end{eqnarray}

Because of rounding to the nearest integer, $p_{\mathrm{up}}(a,t)$ is
only calculated to accuracy $\pm 2^{-\beta}$. Thus, the number of
bitstrings encoding $\includegraphics[width=0.25in]{udsmall.eps}$ can
exceed the number of bitstrings encoding
$\includegraphics[width=0.25in]{dusmall.eps}$ by as much as $\sim
2^\beta$ or vice versa. To achieve unitarity we define 
$U_\sigma$ to act as the identity on these excess bitstrings. We
therefore refer to these strings as ``stuck''.

We will now show that, because these stuck bitstrings form at most a
$\sim 2^{-\beta}$ fraction of all the $2^{2\beta}$ encodings on which
$U_\sigma$ acts, the error introduced by them is negligible, provided
$\beta = \Omega(\log n)$. We can divide the set of bitstrings
$\{0,1\}^{n\beta}$ into those that are stuck and those that are
unstuck. By unitarity, no $U_{\mathrm{pm}}(\sigma_i)$ operator can
ever transform an unstuck string into a stuck string or vice versa. 

The total error introduced by the stuck strings is
\[
E_{\mathrm{stuck}} = \left| \frac{1}{2^{n\beta}} \sum_{x \in
  \{0,1\}^{n\beta}} \left[ \bra{x} U_{\mathrm{pm}}(b) \ket{x} -
  \bra{\eta_h(x)} \rho_{n,k,h}(b) \ket{\eta_h(x)} \right] \right|
\]
For unstuck strings, $\bra{x}
  U_{\mathrm{pm}}(b) \ket{x} = \bra{\eta_h(x)} \rho_{n,k,h}(b)
  \ket{\eta_h(x)}$, thus
\[
E_{\mathrm{stuck}} = \left| \frac{1}{2^{n\beta}} \sum_{x \in S} 
  \left[ \bra{x} U_{\mathrm{pm}}(b) \ket{x} - \bra{\eta_h(x)}
  \rho_{n,k,h}(b) \ket{\eta_h(x)} \right] \right|,
\]
where $S$ is the set of stuck strings. By the triangle inequality
\[
E_{\mathrm{stuck}} \leq \frac{1}{2^{n\beta}} \sum_{x \in S} \left(
  \left| \bra{x} U_{\mathrm{pm}}(b) \ket{x}\right| + \left|
  \bra{\eta_h(x)} \rho_{n,k,h}(b) \ket{\eta_h(x)} \right| \right).
\]
By unitarity these matrix elements have at most unit magnitude, so
\[
E_{\mathrm{stuck}} \leq \frac{1}{2^{n\beta}} \sum_{x \in S} 2.
\]
Thus $E_{\mathrm{stuck}}$ is at most twice the fraction of strings in
$\{0,1\}^{n\beta}$ that are stuck. For each $i=1,2,\ldots,n+1$, the
pair of registers $i$ and $(i+1)$ has probability approximately
$2^{-\beta}$ of being stuck. Thus the fraction of bitstrings in which
at least one pair is stuck is approximately
\[
1-(1-2^{-\beta})^n.
\]
By choosing $\beta = \Omega(\log n)$ we can thus ensure that
$E_{\mathrm{stuck}}$ is polynomially small.

The total error $E$ in the estimate of the normalized trace is
\[
E \leq E_{\mathrm{round}} + E_{\mathrm{stuck}}.
\]
We can see this as follows.
\begin{eqnarray*}
E & = &\left| \frac{1}{|\Omega_{n,k,h}|} \sum_{\omega \in \Omega_{n,k,h}}
\bra{\omega} \rho_{n,k,h}(b) \ket{\omega} - \frac{1}{2^{n\beta}}
\sum_{x \in \{0,1\}^{n\beta}} \bra{x} U_{\mathrm{pm}}(b) \ket{x}
\right| \\
& = &\left| \frac{1}{|\Omega_{n,k,h}|} \sum_{\omega \in \Omega_{n,k,h}}
\bra{\omega} \rho_{n,k,h}(b) \ket{\omega} - \frac{1}{2^{n\beta}}
\sum_{x \in \{0,1\}^{n\beta}} \bra{\eta_h(x)} \rho_{n,k,h}(b)
\ket{\eta_h(x)} \right.\\
& & \left. + \frac{1}{2^{n\beta}} \sum_{x \in \{0,1\}^{n\beta}}
\bra{\eta_h(x)} \rho_{n,k,h}(b) \ket{\eta_h(x)} - \frac{1}{2^{n\beta}}
\sum_{x \in \{0,1\}^{n\beta}} \bra{x} U_{\mathrm{pm}}(b) \ket{x} \right|
\end{eqnarray*}
(We have added and subtracted $\frac{1}{2^{n\beta}}
\sum_{x \in \{0,1\}^{n\beta}} \bra{\eta_h(x)} \rho_{n,k,h}(b) \ket{\eta_h(x)}$,
leaving the total unchanged.) Applying the triangle inequality we obtain
\begin{eqnarray*}
E & \leq & \left| \frac{1}{|\Omega_{n,k,h}|} \sum_{\omega \in \Omega_{n,k,h}}
\bra{\omega} \rho_{n,k,h}(b) \ket{\omega} - \frac{1}{2^{n\beta}}
\sum_{x \in \{0,1\}^{n\beta}} \bra{\eta_h(x)} \rho_{n,k,h}(b)
\ket{\eta_h(x)} \right| \\
& & + \left| \frac{1}{2^{n\beta}} \sum_{x \in \{0,1\}^{n\beta}}
\bra{\eta_h(x)} \rho_{n,k,h}(b) \ket{\eta_h(x)} - \frac{1}{2^{n\beta}}
\sum_{x \in \{0,1\}^{n\beta}} \bra{x} U_{\mathrm{pm}}(b) \ket{x}
\right|.
\end{eqnarray*}
The first term is recognizable as $E_{\mathrm{round}}$, and the second
term is recognizable as $E_{\mathrm{stuck}}$, thus we are done.

Now that we know how to estimate the normalized trace,
\[
\frac{1}{|\Omega_{n,k,h}|} \tr [\rho_{n,k,h}(b)]
\]
for each $h$, we can do weighted classical sampling over $h$ to obtain
the Markov trace, as described in section \ref{pathmodel}. Lastly,
in accordance with equation \ref{Jones_trace}, we multiply by the
easily computed prefactor 
\[
(-ie^{i \pi/2k})^{3w(\vec{L})} (-2\cos(\pi/k))^{n-1}
\]
to obtain an estimate of the Jones polynomial. 

\section{HOMFLY polynomials}
\label{HOMFLY}

As discussed in section \ref{intro}, the Jones polynomial is equivalent
to a special case of a more general knot invariant called the
single-variable HOMFLY polynomial. Let $\vec{L}$ be the trace closure of a
braid $b \in B_n$. To make $\vec{L}$ an oriented link, every strand of the
braid is oriented downward. The single-variable HOMFLY polynomial is
\begin{equation}
\label{HOMFLY_trace}
H_{\vec{L}}^{(r)}(e^{i 2 \pi/k}) = \left( \frac{\sin(\pi r/k)}{\sin(\pi/k)}
  \right)^{n-1} e^{-i (r+1) e(b) \pi/k}
  \widetilde{\tr} (\pi_{n,k,r}(b)) 
\end{equation}
where $\pi_{n,k,r}$ is the Jones-Wenzl representation of $B_n$,
$\widetilde{\tr}$ indicates its Markov trace (to be defined
shortly), and $e(b)$ is the sum of the exponents appearing in $b$
when written in terms of the generators
$\sigma_1,\ldots,\sigma_{n-1}$. Thus, $e(b)$ is minus the writhe of $\vec{L}$.
For each $n$ and $k$, the Jones-Wenzl representation is a unitary
representation of the group $B_n$, whose dimension is exponential in
$n$. In section \ref{HOMFLY_algorithm}, we will describe how to
efficiently implement this unitary representation with quantum
circuits, thereby allowing the efficient estimation of single-variable
HOMFLY polynomials using one clean qubit. In the present section we
will first describe the Jones-Wenzl representation and its Markov
trace. Our presentation closely\footnote{However, for consistency with
  \cite{Aharonov1, Shor_Jordan}, we use $k$ and $r$ to represent the
  parameters called $l$ and $k$, respectively, in \cite{Wocjan_Yard}.}
 \ follows that of \cite{Wocjan_Yard}.

\begin{figure}
\begin{center}
\includegraphics[width=0.4\textwidth]{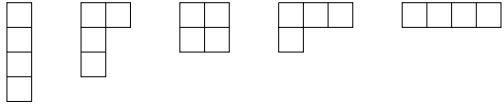}
\caption{\label{Young_diagrams} The Young diagrams with four
  boxes. They correspond to the partitions of the number four.}
\end{center}
\end{figure}

\begin{figure}
\begin{center}
\includegraphics[width=0.7\textwidth]{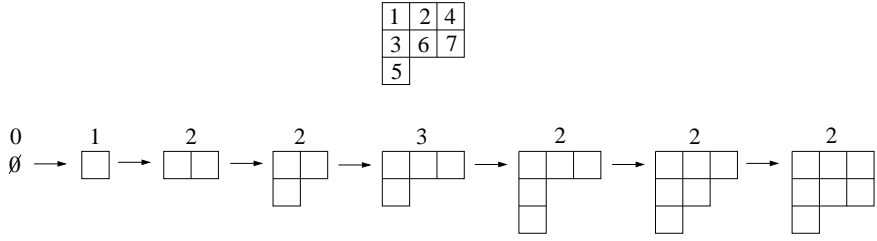}
\caption{\label{Young_tableau} Above we show an example of a standard
  Young tableau, and beneath it the corresponding sequence of Young
  diagrams. Above each Young diagram is listed the number of boxes in
  the first row minus the number of boxes in the third row. (In some
  diagrams the number of boxes in the third row is zero). The maximum
  value taken by this difference is three. Thus, the standard Young
  tableau shown is a member of $T_{7,k,3}$ for $k=6,7,8,\ldots$, but
  not for $k=1,2,3,4,5$.}
\end{center}
\end{figure}

The Jones-Wenzl representation of $B_n$, the braid group of
$n$-strands, is formulated in terms of standard Young tableaux of $n$
boxes. For any $n$, the Young \emph{diagrams} are all the possible
partitions of $n$ boxes into rows, where the rows are arranged in
descending order of length. All the Young diagrams for $n=4$ are
illustrated in figure \ref{Young_diagrams}. For a given Young diagram $\lambda$
the corresponding standard Young \emph{tableaux} are all the
numberings of boxes so that if we started with no boxes, and added
boxes in this order, the configuration would be a valid Young diagram
after every step. An example is shown in figure
\ref{Young_tableau}. 

For the reader intrigued by the appearance of Young tableaux we make
the following aside. Young tableaux were originally introduced to
construct representations of the symmetric group $S_n$ (\emph{cf.}
\cite{Boerner}). $B_n$ is closely related to $S_n$; the latter is
obtained from the former by adding the relation $\sigma_i^2 =
\id$. Any representation $\rho$ of the symmetric group must satisfy
$\rho(\sigma_i)^2 = \id$. By deforming this relation to
$\rho(\sigma_i)^2 = (-t^{-3/4} + t^{1/4}) \rho(\sigma_i) + t^{-1/2}
\id$ we obtain the path model representation of $B_n$. (The
correspondence between paths and standard Young tableaux and the
relationship between the path model and Jones-Wenzl representations
are explained in the appendix.) This type of deformation appears
frequently in mathematics and is referred to as a quantum deformation
or $q$-deformation. In the limit $t \to 1$ we recover a representation
of $S_n$. The origin of the term quantum deformation is the
commutation relation $pq - qp = i \hbar \id$ among the position and
momentum operators in quantum mechanics. This is a deformation of the
classical commutation relation $pq - qp = 0$.

We now describe in detail the Jones-Wenzl representation of $B_n$. Let
$T_{n,k,r}$ be the set of standard Young tableaux of $n$ boxes and
at most $r$ rows, such that after every step, the configuration is not only a
valid Young diagram, but also has the property that the number of
boxes in the first row minus the number of boxes in the $r\th$ row is
at most $k-r$. Let $\mathcal{W}_{n,k,r}$ be the formal span of
$T_{n,k,r}$. For given $n,k,r$, the Jones-Wenzl representation is a
group homomorphism $\pi_{n,k,r}:B_n \to U(\mathcal{W}_{n,k,r})$ from
the braid group $B_n$ to the group of unitary transformations on the
vector space $\mathcal{W}_{n,k,r}$.

The elementary crossings $\sigma_1,\ldots,\sigma_{n-1}$ generate the
braid group $B_n$. Thus, to specify the representation $\pi_{n,k,r}$
it suffices to specify the representations of these crossings \\
$\pi_{n,k,r}(\sigma_1),\ldots,\pi_{n,k,r}(\sigma_{n-1})$, as is done
by the following rule. For any $\Lambda \in T_{n,k,r}$:
\begin{equation}
\label{rule}
\pi_{n,k,r}(\sigma_i) \Lambda = -e^{i \pi(1-d_i(\Lambda))/k}
\frac{\sin(\pi/k)}{\sin(\pi d_i(\Lambda)/k)} \Lambda - e^{i \pi/k}
\sqrt{1-\frac{\sin^2(\pi/k)}{\sin^2(\pi d_i(\Lambda)/k)}} \Lambda',
\end{equation}
where $\Lambda'$ is the Young tableau obtained from $\Lambda$ by
swapping boxes $i$ and $i+1$, and $d_i(\Lambda)$ is the ``axial''
distance from box $i$ to box $i+1$ in $\Lambda$. That is, if box $i$
appears in row $r_i(\Lambda)$ and column $c_i(\Lambda)$ and box $i+1$
appears in row $r_{i+1}(\Lambda)$ and column $c_{i+1}(\Lambda)$ then
\begin{equation}
\label{axial}
d_i(\Lambda) = c_i(\Lambda) - c_{i+1}(\Lambda) - 
\left( r_i(\Lambda) - r_{i+1}(\Lambda) \right).
\end{equation} 
For some $\Lambda \in T_{n,k,r}$, the Young tableau $\Lambda'$
obtained by swapping boxes $i$ and $i+1$ is not contained in
$T_{n,k,r}$. However, one can verify that in such cases, the
coefficient $\sqrt{1-\frac{\sin^2(\pi/k)}{\sin^2(\pi
    d_i(\Lambda)/k)}}$ is always zero. Thus, equation \ref{rule}
defines a linear transformation strictly within
$\mathcal{W}_{n,k,r}$.

\begin{figure}
\begin{center}
\includegraphics[width=0.13\textwidth]{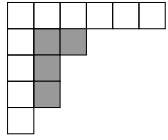}
\caption{\label{hooklength} In the Young diagram shown above, the hook
length of the box at position (2,2) is four. In general the hooklength
of a box is the number of boxes in the ``hook'' that includes the box
itself, all the boxes to the right of it in the same row, and all the
boxes below it in the same column.}
\end{center}
\end{figure}

By swapping boxes, one never changes the shape of a Young
tableau. Thus, the Jones-Wenzl representation is reducible, with
invariant subspaces corresponding to different Young diagrams. The
Markov trace is the following weighted sum of the traces over these
subspaces.
\begin{equation}
\label{Markov_HOMFLY}
\widetilde{\tr}(\pi_{n,k,r}(b)) = \sum_\lambda S^{(\lambda)}_{k,r}
\tr ( \pi_{n,k,r}^{(\lambda)}(b) ),
\end{equation}
where $\pi_{n,k,r}^{(\lambda)}$ is the Jones-Wenzl representation on
the subspace corresponding to Young diagram $\lambda$, $\tr$ denotes
the ordinary matrix trace, and the sum is over all Young diagrams of
$n$ boxes and at most $r$ rows such that the number of boxes in the
top row minus the number of boxes in the $r\th$ row is at most
$k-r$. The weights $S^{(\lambda)}_{k,r}$ are given by
\begin{equation}
\label{weights}
S^{(\lambda)}_{k,r} = \left( \frac{\sin(\pi/k)}{\sin(\pi r/k)}
\right)^n \prod_{(i,j) \in \lambda} \frac{\sin(\pi (j-i+r)/k)}{\sin(\pi
  h_{i,j}(\lambda)/k)},
\end{equation}
where the product is over all (row,column) coordinates in the Young
diagram $\lambda$, and $h_{i,j}(\lambda)$ is the ``hook length'' of
the box at row $i$, column $j$. That is, $h_{i,j}(\lambda)$ is the
number of boxes to the right of box $(i,j)$ in row $i$ plus the number
of boxes below box $(i,j)$ in column $j$, plus 1. This is illustrated
in figure \ref{hooklength}.

\section{Algorithm for HOMFLY polynomials}
\label{HOMFLY_algorithm}

Because of the close relationship between the path model
representation and the Jones-Wenzl representation, the one clean qubit
algorithm for estimating the single-variable HOMFLY polynomial of the
trace closure of a braid is a fairly direct generalization of the
Jones polynomial algorithm of sections \ref{encoding} and
\ref{implementation}. For any fixed $k$ and $r$ the runtime of the
algorithm scales polynomially with $n$. However, we do not have
polynomial scaling with $r$. 

We need an encoding that maps bitstrings to standard Young
tableaux. Let $T_{n,k,r}^{(\lambda)}$ be the set of Young tableaux in
$T_{n,k,r}$ compatible with Young diagram $\lambda$. For each
$\lambda$ we introduce
\[
\nu_\lambda: \{ 0,1 \}^{n\beta} \to T_{n,k,r}^{(\lambda)}.
\]
In order to get a uniformly weighted trace, we must construct a
$\nu_\lambda$ with the property that
\begin{equation}
\label{equiprobable}
| \nu_\lambda^{-1}(\Lambda)| \simeq
  \frac{2^{n\beta}}{|T_{n,k,r}^{(\lambda)}|} 
\end{equation}
for each $\Lambda \in T_{n,k,r}^{(\lambda)}$. To design a mapping
$\nu_\lambda$ satisfying equation \ref{equiprobable}, we think in
terms of a classical randomized algorithm for uniformly sampling from
$T_{n,k,r}^{(\lambda)}$ using $n\beta$ random bits. The algorithm works
similarly to the algorithm described in section \ref{encoding} for
sampling from the paths $\Omega_{n,k,h}$. The main difference is that
at each step in a path, one has at most two choices: step up or step
down, whereas at each step in the sequence corresponding to a Young
tableau of $r$ rows, one can have as many as $r$ choices: add a box to
any row. To ensure a uniform sampling from $T_{n,k,r}^{(\lambda)}$, we
must probabilistically make this choice as follows. After choosing the
first $t < n$ steps we have a Young tableau $\Lambda_t \in
T_{t,k,r}$. Let $R_k^{(\lambda)}(\Lambda_t)$ be the number of Young
tableaux in $T_{n,k,r}^{(\lambda)}$ obtainable by starting with
$\Lambda_t$ and adding the remaining $n-t$ boxes. Let $\Lambda_t^j$
be the Young tableau obtained from $\Lambda_t$ by adding the next box
to row $j$. At each step we must add a box to row $j$ with probability
\begin{equation}
\label{pbox}
p_j^{(\lambda)}(\Lambda_t) =
\frac{R_k^{(\lambda)}(\Lambda_t^j)}{R_k^{(\lambda)}(\Lambda_t)}.
\end{equation}
Note that there are two cases where $R_k^{(\lambda)}(\Lambda_t^j)=0$.
The first is when $j=1$, and by adding this last box to the top row we
violate the condition that the number of boxes in the top row of
$\Lambda_t^j$ minus the number of boxes in the bottom row of
$\Lambda_t^j$ must be at most $r-k$. The second case is when
$\Lambda_t$ has an equal number of boxes in rows $j$ and $j-1$. Thus
by adding a box to row $j$ we obtain an invalid Young diagram.

To generate a random element of $T_{n,k,r}^{(\lambda)}$ we use $n$
registers of $\beta$ random bits. We think of these registers as encoding
numbers $r_1,\ldots,r_n$ in the range $0,1,\ldots,2^\beta-1$. Let $F$ be
the cumulative distribution function
\begin{equation}
\label{cumulative}
F_j(\Lambda_t,\lambda) = \sum_{i=1}^j p_i^{(\lambda)}(\Lambda_t),
\end{equation}
with $F_0(\Lambda_t,\lambda) = 0$. The $(t+1)\th$ box is added to row $j$
if and only if 
\[
\lceil F_{j-1}(\Lambda_t,\lambda) 2^\beta \rceil \leq r_t <  \lceil
F_j(\Lambda_t, \lambda) 2^\beta \rceil. 
\]
By doing this, we choose which row to add each box to approximately
according to equation \ref{pbox}. By essentially the same argument
given in section \ref{implementation}, it suffices to use
probabilities $p_j^{(\lambda)}(\Lambda_t)$ accurate to within 
$\pm \frac{1}{\mathrm{poly}(n)}$. Hence, we can again choose 
$\beta = \mathcal{O}(\log n)$.

For each $\sigma_i \in B_n$ we show how to efficiently implement a
quantum circuit $U_{\mathrm{JW}}(\sigma_i)$ such that for almost all
$x,y \in \{0,1\}^{n\beta}$,
\begin{equation}
\label{encoded_unitary}
\bra{x} U_{\mathrm{JW}}(\sigma_i) \ket{y} \simeq \bra{\nu_\lambda(x)}
\pi_{n,k,r}^{(\lambda)}(\sigma_i) \ket{\nu_\lambda(y)}.
\end{equation}
By concatenating these circuits, we can efficiently implement the
Jones-Wenzl representation of any braid of polynomially many
crossings. Then, by using the one clean qubit algorithm for trace
estimation, we can approximate the HOMFLY polynomial of the trace
closure of the braid.

$\pi_{n,k,r}^{(\lambda)}(\sigma_i)$ transforms only boxes $i$ and
$i+1$. By the definition of $\nu_\lambda$ the location of these two
boxes is encoded in the $i\th$ and $(i+1)\th$ register of $\beta$ qubits
each. Thus, $U_{\mathrm{JW}}(\sigma_i)$ transforms only these two
registers. By equation \ref{rule} it is apparent that the
transformation performed on these two registers depends on the axial
distance between the boxes they describe. Less obviously, the
transformation depends on the cutoffs 
\[
\lceil F_j(\Lambda_i,\lambda) 2^\beta \rceil, \quad \lceil
F_{j'}(\Lambda_{i+1},\lambda) 2^\beta \rceil 
\]
for certain relevant $(j,j')$. This is because these cutoffs determine the
encoding $\nu_\lambda$ between Young tableaux and bitstrings.

The axial distance and the cutoffs are encoded in the preceding
$(i-1)$ $\beta$-qubit registers. We'll show how to extract the relevant
information into logarithmically many ancilla qubits, so that the
transformation $U_{\mathrm{JW}}(\sigma_i)$ can be implemented by a
quantum circuit acting on only logarithmically many qubits. By the
general construction of \cite{Nielsen_Chuang_twolevel}, any unitary on
logarithmically many qubits can be implemented using polynomially many
quantum gates.

\begin{figure}
\begin{center}
\includegraphics[width=0.37\textwidth]{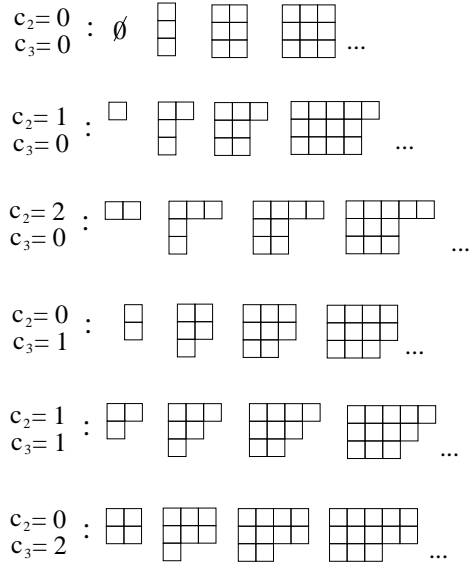}
\caption{\label{profiles} As an example we use $r=3$, $k=2$. We display
the corresponding Young diagrams for each allowed profile
$(c_2,c_3)$, where $c_2$ is the ``overhang'' of the top row over the
second row, and $c_3$ is the overhang of the second row over the
bottom row. As we add boxes, the length of these overhangs
changes. Thus, each Young tableau in $T_{n,2,3}$ uniquely corresponds
to an $n$-step walk on the six allowed profiles.}
\end{center}
\end{figure}

Rather than directly computing cutoffs and axial distances, we'll work
in terms of other quantities which are easier to extract from the
first $(i-1)$ registers. Recall that a Young tableau can be thought of
as a sequence of steps by which to build a final Young diagram, adding
one box at a time. Let $b_j(t)$ be the number of boxes in row $j$
after $t$ steps. $b_1(t),b_2(t),\ldots,b_r(t)$ completely describe the
Young diagram of step $t$. We can do a change of variables, defining
\begin{eqnarray*}
c_1(t) & = & b_1(t) + b_2(t) + \ldots + b_r(t) = t\\
c_2(t) & = & b_1(t)-b_2(t) \\
c_3(t) & = & b_2(t)-b_3(t) \\
& \vdots & \\
c_r(t) & =  & b_{r-1}(t)-b_r(t). \\
\end{eqnarray*}
The $(r-1)$-tuple
\[
\vec{c}(t) = (c_2(t),c_3(t),\ldots,c_r(t))
\]
defines the ``profile'' of the Young tableau, as illustrated in figure
\ref{profiles}. These profiles are higher dimensional analogues to the
rungs in the path model. The restriction to $k$ rungs is here replaced
with the restriction to profiles in which $c_2+c_3+\ldots+c_r \leq
k-r$. The Jones-Wenzl representation acts on the space of Young
tableaux which correspond to walks on these profiles, just as the path
model representation acts on the space of paths which correspond to
walks on the rungs.

We'll next show how to extract $\vec{c}(i-1)$ into $\mathcal{O}((r-1)
\log n)$ clean ancilla qubits. Once we do this, we can implement
$U_{JW}(\sigma_i)$ because its action on the $i\th$ and $(i+1)\th$
registers is completely determined by $\vec{c}(i-1)$. In order to
compute $\vec{c}(i-1)$ we need to know the cutoffs $\lceil
F_j(\Lambda_t,\lambda) 2^\beta \rceil$ for all $t < i-1$ and all
relevant $j$. The key thing to notice about $\lceil
F_j(\Lambda_t,\lambda) 2^\beta \rceil$ is that it depends only on $t$,
$j$, $\lambda$ and the profile of $\Lambda_t$, not on any of
$\Lambda_t$'s internal details. As a result, for any fixed\footnote{As
  described at the end of this section, we classically sample over
  $\lambda$. Thus, each time we run the one clean qubit computer
  $\lambda$ has some random fixed value.} \ $r$, $k$, and $\lambda$,
there are only polynomially many cutoffs we need to compute, which we
can see as follows. $c_2,c_3,\ldots,c_r$ are all upper bounded by
$k-r$, thus $(k-r)^{r-1}$ provides a loose upper bound on the number
of allowed values of $\vec{c}(t)$. $t$ runs from 1 to $n$ and $j$ runs
from 1 to $r$. Thus the total number of cutoffs we need to compute is
upper bounded by $r n (k-r)^{r-1}$, which is exponential in $r$, but
for any fixed $r$ is polynomial in $n$. Thus we can classically
precompute all of the necessary cutoffs and store them in a classical
lookup table.

\begin{figure}
\begin{center}
\includegraphics[width=0.63\textwidth]{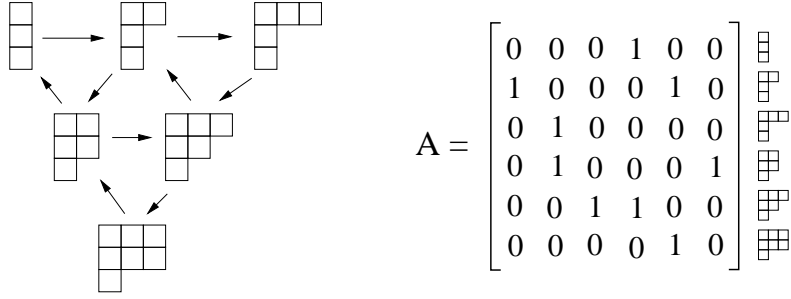}
\caption{\label{digraph} Continuing the example in figure
  \ref{profiles} we choose $r=3$, $k=2$. We display a
  representative Young diagram for each allowed profile. The arrows
  represent the allowed transitions between these profiles by adding
  one box. $A$ is the adjacency matrix of this directed graph. The
  arrows to the right represent the addition of a box to the top
  row, the arrows diagonally downward represent the addition of a box
  to the middle row, and the arrows diagonally upward represent the
  addition of a box to the bottom row. After adding a box to the
  bottom row we omit the leftmost complete column, as per the notation
  of figure \ref{profiles}.}
\end{center}
\end{figure}

We will classically compute, for each of the allowed profiles
of $\vec{c}(t)$, and each $j$ and $t$, the corresponding cutoff
\begin{equation}
\label{cutoff}
\lceil F_j(\vec{c}(t),\lambda, t) 2^\beta \rceil.
\end{equation}
To do this, we imagine a directed graph with vertices
corresponding to the allowed profiles. An edge leads from profile
$\vec{a}$ to profile $\vec{b}$ if $\vec{b}$ can be obtained from
$\vec{a}$ by adding one box. This is illustrated in figure
\ref{digraph}. If we take the adjacency matrix $A$ of this graph, and
raise it to power $s$, the matrix elements are equal to the number of
ways of getting from one profile to another using $s$ steps. In this
way, we can obtain the value of $R_k^{(\lambda)}(\Lambda_t)$
needed in equation \ref{pbox}. This is the number of ways to get from
$\vec{c}(t)$ to $\vec{c}(n)$ (the profile of $\lambda$) using $n-t$
steps. Similarly, we can obtain $R_k^{(\lambda)}(\Lambda_t^j)$, which
is the number of ways of getting to $\vec{c}(n)$ by starting with the
profile of $\Lambda_t^j$ and making $n-t-1$ steps. Thus, after
computing the relevant powers of $A$, we can then efficiently
compute each $\lceil F_j(\vec{c}(t),\lambda, t) 2^\beta \rceil$ using
equations \ref{pbox} and \ref{cumulative}. 

Given our table of cutoffs, the procedure for computing $\vec{c}(t)$
is a simple iteration. Suppose we know $\vec{c}(t-1)$. To obtain
$\vec{c}(t)$ we compare the $t\th$ register of $\beta$ qubits to the
relevant cutoffs 
\[
\lceil F_1(\vec{c}(i-1),\lambda, i-1) 2^\beta \rceil, \ldots, \lceil
F_r(\vec{c}(i-1),\lambda,i-1) 2^\beta \rceil
\]
to determine which row the $t\th$ box is added to. If the $t\th$ box
is added to row $j$, then we decrement $c_j$ (unless $j=1$) and
increment $c_{j+1}$.

The $i\th$ and $(i+1)\th$ registers together with the ancilla qubits
containing $\vec{c}(i-1)$ encode the locations of boxes $i$ and
$i+1$. Thus, we can perform the transformation
$U_{\mathrm{JW}}(\sigma_i)$, as specified by equations \ref{rule} and
\ref{encoded_unitary} using a quantum circuit that acts only on these
qubits. More specifically, this quantum circuit performs a unitary
transformation on the $i\th$ and $(i+1)\th$ registers that depends on the
content of the ancilla qubits. The ancilla qubits themselves are not
transformed.

The unitary transformation performed on the $i\th$ and $(i+1)\th$
registers is one which rotates between the encodings of a pair
standard Young tableaux which differ by having boxes $i$ and $(i+1)$
swapped. $\nu_\lambda$ is not injective, but the number of bitstrings
which encode these two tableaux are approximately equal. We illustrate
this with an example. Suppose
$\includegraphics[width=0.4in]{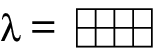}$. Consider the following
pair of standard Young tableaux
\[
\includegraphics[width=0.26\textwidth]{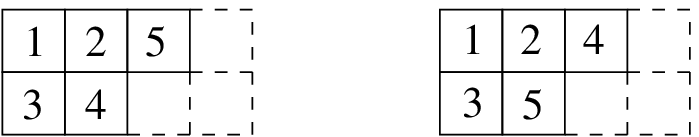}.
\]
The total number of standard Young tableaux of shape $\lambda$ whose
first five boxes appear in the configuration shown at left is the same
as the number of standard Young tableaux of shape $\lambda$ whose
first five boxes appear in the configuration shown at right. This is
because this number depends only on the shape of the dashed
region. Returning to the general case, we see that swapping a pair of
labelled boxes can never change the shape of the dashed region. By the
definition of $\nu_\lambda$, the fraction of the $2^{2 \beta}$
possible bitstrings for registers $i$ and $i+1$ that encode a given
configuration of boxes $i$ and $i+1$ is proportional to the fraction
of Young tableaux of shape $\lambda$ in which the boxes are in that
configuration. Hence, number of bit assignments for registers $i$ and
$i+1$ that encode a given configuration is equal to the number that
encode the configuration in which boxes $i$ and $i+1$ are swapped, up
to rounding. Thus we can always make some canonical matching between
the bitstrings encoding the two configurations. The encoded version of
transformation \ref{rule} is then to unitarily rotate between the
current bitstring and its canonical matching.

In the case of Jones polynomials, we specified the canonical matching
in equation \ref{placevalue}. Here due to greater complexity we do not
specify any formula for the matching. Instead, while computing
all the cutoffs, one can at the same time make arbitrary choices
for the corresponding matchings and write them down. A complete lookup
table of these choices can be stored using polynomially many bits
because $2\beta = \mathcal{O}(\log n)$. Given the choices of
matchings, one can then use equation \ref{rule} to calculate all the
matrix elements of $U_{\mathrm{JW}}(\sigma_i)$. This matrix has
polynomial dimension since it acts only on the two registers
$\beta=\mathcal{O}(\log n)$ qubits each plus the $\mathcal{O}(\log n)$
ancillas encoding $\vec{c}(i-1)$.  It can therefore be implemented by
an efficient quantum circuit using the method of
\cite{Nielsen_Chuang_twolevel}. After performing the unitary
transformation, $\vec{c}(i-1)$ can be uncomputed.

As mentioned above, because of rounding, the number of bitstrings
encoding the swapped and unswapped pair of boxes are not precisely
equal, only approximately equal. Thus our canonical matching will in
general have a small number of unpaired bitstrings encoding one of the
two tableaux. As we did for Jones polynomials we define the unitary
transformation to act as the identity on these excess bitstrings outside
of the matching. By an analysis essentially identical to that in
section \ref{implementation} one can see that choosing $\beta$
logarithmic in the number of crossings suffices to ensure that these
unmatched bitstrings form a small enough fraction so that their effect
on the trace of the circuit is negligible.

By the above procedure we can construct an efficient quantum circuit
for $U_{\mathrm{JW}}(\sigma_i)$ satisfying equation \ref{encoded_unitary} for
any $i$ and any $\lambda$. By concatenating these, we can thus obtain
a quantum circuit for $U_{\mathrm{JW}}(b)$ for any $b \in B_n$ of polynomially
many crossings. If $\vec{L}$ is the link obtained by taking the
trace closure of $b$ with each strand oriented downward then the
corresponding HOMFLY polynomial is given by the Markov trace
\[
H_{\vec{L}}^{(r)}(e^{i 2 \pi/k}) = \left( \frac{\sin(\pi k/r)}{\sin(\pi/k)}
\right)^{n-1} e^{-i \pi (r+1) e(b)/k} \sum_\lambda
S_{k,r}^{(\lambda)} \tr \left( \pi^{(\lambda)}_{n,k,r}(b) \right).
\]
For any $\lambda$ we can estimate the normalized trace of
$U_{\mathrm{JW}}(b)$ to polynomial precision using the 
standard one clean qubit algorithm for trace estimation. Thus, we can
estimate the HOMFLY polynomial by classically sampling from the
possible Young diagrams $\lambda$ according to the distribution
\[
p(\lambda) = \frac{S_{k,r}^{(\lambda)}
  |T_{n,k,r}^{(\lambda)}|}{\sum_{\lambda'} S_{k,r}^{(\lambda')} 
|T_{n,k,r}^{(\lambda')}|} 
\]
and estimating the corresponding normalized trace 
\[
\frac{1}{|T_{n,k,r}^{(\lambda)}|} \tr(U_{\mathrm{JW}}(b))
\]
for each $\lambda$ sampled.

To do this we need to compute the values of $S_{k,r}^{(\lambda)}$ and
$|T_{n,k,r}^{(\lambda)}|$ for each allowed $\lambda$. It is not hard
to see that the allowed $n$-box Young diagrams are in bijective
correspondence with the allowed profiles. Thus for fixed $k$ and $r$,
the number of values of $S_{k,r}^{(\lambda)}$ we need to compute is
independent of $n$. It is clear by equation \ref{weights} that each
$S_{k,r}^{(\lambda)}$ can be classically computed in polynomial
time. Similarly, for fixed $k$ and $r$, there are only
$\mathrm{poly}(n)$ different values of $|T_{n,k,r}^{(\lambda)}|$ to
compute. $|T_{n,k,r}^{(\lambda)}| = R_k^{(\lambda)}(\emptyset)$,
thus each $|T_{n,k,r}^{(\lambda)}|$ can be computed in polynomial time
using the algorithm for computing $R_k^{(\lambda)}$ described
earlier.

\section{Conclusion}

In this paper we have shown that one clean qubit computers can in
polynomial time obtain additive approximations to the Jones and HOMFLY
polynomials of the trace closure of braids at arbitrary roots of
unity. This generalizes the result of \cite{Shor_Jordan} which showed
that one clean qubit computers can efficiently approximate the Jones
polynomial of the trace closure of braids at the fifth root of
unity. In \cite{Shor_Jordan} it was also shown that this problem is
DQC1-complete. The completeness proof is based on the fact that the image of
the path model representation $\rho_{n,5}:B_n \to
U(\mathcal{V}_{n,5})$ modulo global phase is dense in
$SU(\mathcal{V}_{n,5})$. By the results of \cite{Freedman3}, this
density result holds also for all $k$ other than 1,2,3,4, and 6, and
similar density results hold for the Jones-Wenzl representation. Thus
it is natural to conjecture that DQC1-completeness extends to Jones
polynomials beyond $k=5$ and to HOMFLY polynomials. DQC1-completeness
would imply that the additive approximations achieved by the
algorithms here cannot be achieved in polynomial time by classical
computers unless DQC1 $\subseteq$ P. Such completeness questions
provide a promising direction for further research.

Another direction is to generalize the algorithm even further. For
evaluating the Jones polynomial when $t$ is not a root of unity, the
relevant representation of the braid group is nonunitary. In
\cite{Aharonov3}, Aharonov \emph{et al.} give a general quantum
algorithm to approximate Jones polynomials at all values of $t$ and to
evaluate Tutte polynomials. They achieve this by interacting the
computational qubits with an ``environment'' of ancilla qubits thereby
inducing nonunitary dynamics on the computational qubits. It would be
interesting to see whether similar techniques can be carried over to
the one clean qubit model.

\section{Acknowledgements}

We thank Peter Shor for useful discussions. During the research and
writing of this paper SJ was at Center for Theoretical physics at MIT,
the Digital Materials Laboratory at RIKEN, and the Institute for
Quantum Information at Caltech. SJ thanks these institutions as well
as the Army Research Office (ARO), the Disruptive Technology Office
(DTO), the Department of Energy (DOE), Franco Nori and Sahel Ashab at
RIKEN, and John Preskill at Caltech. PW gratefully acknowledges
support from NSF grants CCF-0726771 and CCF-0746600. PW would like to
thank Eddie Farhi's group for their hospitality and the W. M. Keck
Foundation for partial support.

\appendix

\section{ Jones Polynomials from HOMFLY polynomials }

\begin{figure}
\begin{center}
\includegraphics[width=0.45\textwidth]{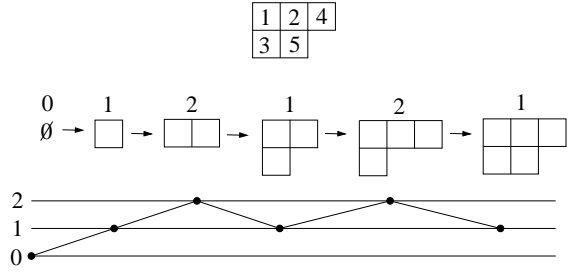}
\caption{\label{correspondence} For the special case of two rows, the
  Young tableaux of $n$ boxes become equivalent to paths of $n$
  steps. Adding a box to the top row corresponds to a step up, and adding a box
  to the bottom row corresponds to a step down.}
\end{center}
\end{figure}

As shown in figure \ref{Young_tableau}, a Young tableau corresponds to
a process by which a Young diagram is built up by adding one box at a
time. If $r=2$ then the Young diagram has two rows (although at some
steps the second row may be empty). This process can therefore be
completely described by listing the difference between the number of
boxes in the first and second rows at each step. The values of this
difference correspond to the rungs of the ladder in the path model, as
illustrated in figure \ref{correspondence}. The values appearing in
the path model representation, as defined in section \ref{pathmodel},
can be rewritten as follows.
\[
\begin{array}{rclcl}
\displaystyle a_l & = & \displaystyle c_{-l} & = & \displaystyle i
  e^{-i \pi \frac{2l+1}{2k}} \frac{\sin(\pi/k)}{\sin(\pi l/k)}
  \vspace{6pt} \\
\displaystyle b_l & = & \displaystyle d_l & = & \displaystyle i e^{-i
  \pi/2k} \sqrt{1-\left(\frac{\sin(\pi/k)}{\sin(\pi l/k)} \right)^2}
\vspace{6pt} \\ 
\displaystyle e_l & = & \displaystyle f_l & = & \displaystyle i e^{-i
  \pi/2k} 
\end{array}
\]
Thus, comparing the path model representation to equation \ref{rule}
shows that
\begin{equation}
\label{rep_correspondence}
\pi_{n,k,2}(\sigma_i) = i e^{i 3 \pi/2k} \rho_{n,k}(\sigma_i).
\end{equation}

\begin{figure}
\begin{center}
\includegraphics[width=0.8\textwidth]{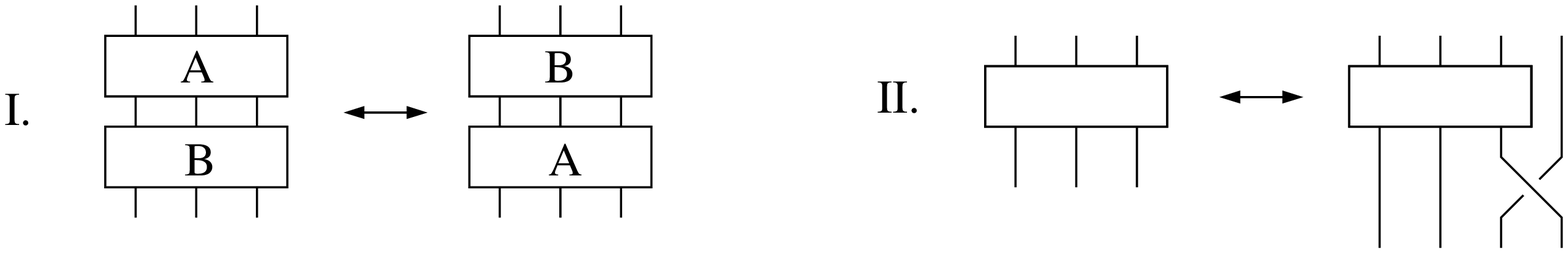}
\caption{\label{Markov_moves} Shown are the two Markov moves. Here the
  boxes $A$ and $B$ represent arbitrary braids. Note that Markov move
  II increases the number of strands by one. }
\end{center}
\end{figure}

As shown in section 5 of \cite{Wocjan_Yard}, the weights in the Markov
trace for the Jones-Wenzl representation simplify substantially in the
case $r=2$. Specifically, the weights $S^{(\lambda)}_{k,r}$ given in
equation \ref{weights} simplify to
\[
S^{(\lambda)}_{k,2} = \frac{\sin(\pi l(\lambda) /k)}{\sin(\pi/k) (2
  \cos(\pi/k))^n}, 
\]
where $l(\lambda)$ is the number of boxes in the top row of $\lambda$
minus the number of boxes in the bottom row of $\lambda$ plus 1. By
the correspondence of figure \ref{correspondence}, $l$ is the
final rung of the corresponding path. The Markov trace of the
Jones-Wenzl representation of the identity braid is 1. Thus
\[
\sum_\lambda |T^{(\lambda)}_{n,k,2}| S^{(\lambda)}_{k,2} = 1,
\]
and so
\[
S^{(\lambda)}_{k,2} = \frac{1}{\sum_{\lambda'} |T^{(\lambda')}_{n,k,2}|
  \sin(\pi l(\lambda') /k)} \sin(\pi l(\lambda) /k),
\]
where the sum over $\lambda'$ is over all Young diagrams of $n$ boxes
and 2 rows such that $l(\lambda') < k$. Comparison with equation
\ref{Markov_Jones} shows that the weighted traces appearing in the Jones and
HOMFLY polynomials are weighted identically in the case $r=2$. This
fact and equation \ref{rep_correspondence} show that for any
braid $b \in B_n$,
\begin{equation}
\label{trace_correspondence}
\widetilde{\tr}(\pi_{n,k,2}(b)) = (i e^{i 3 \pi/2k})^{e(b)} 
\widetilde{\tr}(\rho_{n,k}(b)),
\end{equation}
where $e(b)$ is the sum of the exponents appearing in $b$ when written
in terms of the generators $\sigma_1, \ldots,
\sigma_{n-1}$. Substituting equation \ref{trace_correspondence} into
equation \ref{HOMFLY_trace} and simplifying yields
\[
H_{\vec{L}}^{(2)}(e^{i 2 \pi/k}) = i^{e(b)} (2 \cos(\pi/k))^{n-1} 
e^{-i 3 e(b) \pi/2k} \widetilde{\tr}(\rho_{n,k}(b))
\]
where $\vec{L}$ is the directed link obtained by taking the trace
closure of the braid $b$. $e(b)$ is minus the writhe of
$\vec{L}$. Thus, comparison with equation \ref{Jones_trace} shows
\begin{eqnarray*}
H_{\vec{L}}^{(2)}(e^{i 2 \pi/k}) & = & (-i)^{-2w(\vec{L})} (-1)^{n-1}
V_{\vec{L}}(e^{i 2 \pi/k}) \\
& = & (-1)^{w(\vec{L})+n-1} V_{\vec{L}}(e^{i 2 \pi/k}).
\end{eqnarray*}

The sign discrepancy $(-1)^{w(\vec{L})+n-1}$ is itself a link
invariant, and is therefore inconsequential. To show this we use Markov's
theorem, which states that the oriented link obtained by taking the
trace closure of braid $b_1$ is equivalent to the oriented link obtained
by taking trace closure of braid $b_2$ if and only if $b_1$ can be
transformed into $b_2$ by some finite sequence of the two Markov moves
shown in figure \ref{Markov_moves} (and their inverses).
It is easy to see that the factor $(-1)^{w(\vec{L})+n-1}$ is
invariant under both Markov moves for all braids and is therefore an
invariant of the corresponding trace closures.

\bibliography{generalk}

\end{document}